\pdfoutput=1
\documentclass[showpacs,amsmath,amssymb,aps,pre,twocolumn]{revtex4-1}
\usepackage{graphicx}
\usepackage{dcolumn}
\usepackage{bm}
\usepackage[breaklinks,colorlinks = true,linkcolor = red,urlcolor=blue,citecolor=red]{hyperref}
\usepackage{multirow}
\usepackage{array}
\usepackage{booktabs}
\usepackage{ctable}
\usepackage{ulem}
\usepackage{upgreek}
\usepackage{epsfig}
\usepackage{mathrsfs}
\usepackage{amssymb}
\usepackage{amsbsy}
\usepackage{color}
\usepackage{cancel}
\usepackage{pifont}
\usepackage{marginnote}
\usepackage{float}
\usepackage{verbatim}
\usepackage{subfigure}
\usepackage{bbm, dsfont}
\usepackage{lineno}
\usepackage{amsmath}
\definecolor{dgreen}{rgb}{0,0.7,0}

\def\bea{\begin{eqnarray}}
\def\eea{\end{eqnarray}}

\def\nn{\nonumber}

\makeatletter

\newcommand{\Rmnum}[1]{\expandafter\@slowromancap\romannumeral #1@}
\makeatother

\newcommand{\eref}[1]{Eq.~(\ref{#1})}%
\usepackage[utf8]{inputenc}

\begin{abstract}
We revisit the simple lattice random walk (P\'{o}lya walk) and the Sisyphus random walk in $\mathbb{Z}$, in the presence of random restarts. We use a relatively direct approach namely \textit{First passage under restart for discrete space and time} which was recently developed by us in PRE \textbf{103}, \textit{052129} (2021) and rederive the first passage properties of these walks under the memoryless geometric restart mechanism. Subsequently, we show how our method could be generalized to arbitrary first passage process subject to more complex restart mechanisms such as sharp, Poisson and Zeta distribution where the latter is heavy tailed. We emphasize that the method is very useful to treat such problems both analytically and numerically.
\end{abstract}

\begin{document}

\title{The P\'{o}lya and Sisyphus lattice random walks with resetting -- a first passage under restart approach}

\author{Ofek Lauber Bonomo and Arnab Pal}
\email{richard86arnab@gmail.com}

\affiliation{ School of Chemistry, Raymond and Beverly Sackler Faculty of Exact Sciences \& The Center for Physics and Chemistry of Living Systems \&   The Ratner Center for Single Molecule Science, Tel Aviv University, Tel Aviv 6997801, Israel}


\maketitle
\section{Motivation}
 First passage under restart recently emerged as a useful framework to study arbitrary stochastic processes under arbitrary restart mechanisms \cite{FPUR}. Although the problems with continuous space and time set-ups have been studied in full glory, their discrete counterparts have been largely overlooked. To bridge this void, in a recent work \cite{RW1}, we developed a \textit{First Passage Under Restart} formalism for \textit{discrete space and time processes}. This comprehensive framework can be used to derive exact and useful formulae such as the generating functions for the survival and first passage, the mean first passage time (and higher order moments) etc. in the presence of arbitrary restart distributions. The derivation does not assume any specific form for the underlying first passage process or the restart process, hence can be ubiquitously used in a generic set-up.

 In this note, power of this approach is illustrated using the following examples: P\'{o}lya's lattice random walk on the integer points of the real line, and the one sided and double sided Sisyphus random walk in the presence of geometric restart -- a memoryless distribution. These models were studied earlier and we will refer to these works in proper context. Moving forward, we show our method could be used to accommodate any first passage process that maybe subject to more complex restart distributions (and not necessarily memoryless) namely sharp, Poisson and Zeta. In the conclusions, we sketch out a three-step blue-print on how to use our recipe in a plug-n-play manner either analytically or numerically.

\section{Notations}

 Before we proceed, it will be useful to introduce the notations used throughout the paper. We will use $P_X(x)$, $\langle X \rangle$, $\sigma_X ^2$, and $G_X(z) \equiv \langle z^X \rangle$ to denote, respectively, the probability mass/density function (PMF/PDF), mean/expectation, variance, and the probability generating function (PGF) of a discrete random variable $X$ taking values in the non-negative integers. In addition, the survival function and its corresponding generating function for the underlying process are denoted by $Q_{N}(n)$ and $G_{Q_N}(z)$ respectively. In the presence of restart, generating function for the survival function will be denoted by $G_{Q_{N_R}}(z)$.

\section{Discrete First passage under discrete restart}
\label{FPUR}
Consider a generic discrete step first passage process that starts at the origin and, if allowed to take place without interruptions, ends after a random number of steps $N$. The process is, however, restarted after some random number of steps $R$. Thus, $P_{N}(n)$ and $P_{R}(n)$ are the probability density functions for the first passage and restart process respectively. If the process is completed \textbf{only prior} to the restart, we mark a completion of the event. Otherwise, the process will start from scratch and begin completely anew. This procedure repeats itself until the process reaches completion. Denoting the random completion number of steps of the restarted process by $N_R$, it can be seen that
\bea
N_{R}=\begin{cases}
N & N < R\\
R+N'_{R} & N \geq R, \label{Renewal eq}
\end{cases}
\eea 
where $N'_{R}$ is an independent and identically distributed copy of $N_{R}$. \eref{Renewal eq} is the central renewal equation for first passage step under restart and assumes that after each restart, the memory is erased from the previous trial. To obtain the mean number of steps for the restarted process, we note that \eref{Renewal eq} can be written as $N_R=min\left( N,R \right)+I\left\{ N \geq R\right\} N'_{R}$, where $I\left\{ N  \geq R\right\}$ is an indicator random variable that takes the value 1 when $N \geq R$ and zero otherwise. \textbf{This approach is slightly different than the one used in \cite{RW1} where we assumed that process can complete even when the steps for first passage and restart coincide i.e., $N_R=N$ when $N \leq R$ unlike here [see \eref{Renewal eq} in above].}

\subsection{Mean}

We take expectations on both sides of \eref{Renewal eq} and use the fact that $N$ and $R$ are independent of each other. This results in the following expression for the \textit{mean completion time} under restart
\bea 
\left\langle N_{R}\right\rangle =\frac{\left\langle min\left(N,R\right)\right\rangle }{\text{Pr}\left(N < R\right)}. \label{mean FPUR}
\eea 
The numerator can be computed by noting that the probability $\text{Pr}\left(min\left(N,R\right)>n\right)=\text{Pr}\left(N>n\right)\text{Pr}\left(R>n\right)$ and thus
\bea 
\left\langle min\left(N,R\right)\right\rangle&=&\sum_{n=0}^{\infty}\text{Pr}\left(min\left(N,R\right)>n\right) \nn \\
&=& \sum_{n=0}^{\infty} \left(\sum_{k=n+1}^{\infty}P_N\left(k\right)\right) \left(\sum_{m=n+1}^{\infty}P_R\left(m\right)\right) \nn \\
&=& \sum_{n=0}^{\infty} Q_N(n) Q_R(n)
, \label{Mean min formula}
\eea 
where note that $Q_N(n)$ and $Q_R(n)$ are the survival functions for the first passage and restart processes respectively i.e.,
\begin{align}
    Q_N(n) &=\text{Pr}(N>n), \label{Survival N} \\
    Q_R(n)&=\text{Pr}(R>n) \label{Survival R}.
\end{align}
The denominator in \eref{mean FPUR} can also be computed easily 
\bea
\text{Pr}\left(N < R\right)=\sum_{n=0}^{\infty}P_{N}(n)\sum_{m=n+1}^{\infty}P_{R}(m). \label{denominator}
\eea

\subsection{Generating function for the restarted process}

We now turn our attention to derive the generating function for the restarted process. The probability generating function of the discrete random variable $N_R$ taking values in the non-negative integers ${0,1, ...}$ is defined as
\bea 
G_{N_R}(z) \equiv \left\langle z^{N_R}\right\rangle = \sum_{n=0}^{\infty} P_{N_R}(n)z^{n}, \label{PGF def}
\eea 
where $P_{N_R}(n)$ is the probability mass function of $N_R$. It will now prove useful to introduce the following conditional random variables
\bea 
N_{min}&\equiv & 
\left\{ N|N < R\right\} 
, \label{Tmin main} \\
R_{min}&\equiv &\left\{ R|N \geq R\right\} 
, \label{Rmin main}
\eea 
with their respective densities
\bea 
P_{N_{min}}(n)&=&P_{N}(n)\frac{\sum_{m=n+1}^{\infty}P_{R}(m)}{\text{Pr}\left(N < R\right)}, \label{Nmin mass-main} \\
P_{R_{min}}(n)&=&P_{R}(n)\frac{\sum_{m=n}^{\infty}P_{N}(m)}{\text{Pr}\left(N \geq R\right)}, \label{Rmin mass-main}
\eea 
where $\text{Pr}\left(N \geq R\right)=1-\text{Pr}\left(N < R\right)$.
Using the renewal \eref{Renewal eq} and the new random variables in Eqs.(\ref{Tmin main})-(\ref{Rmin main}), we can write
\begin{align}
    G_{N_{R}}(z)&=&\text{Pr}\left(N < R\right)\left\langle z^{N_{min}}\right\rangle +\text{Pr}\left(N \geq R\right)\left\langle z^{R_{min}+N'_{R}}\right\rangle.
\end{align}
Now using the fact that $N'_{R}$ is an independent
and identically distributed copy of $N_{R}$ in above, we arrive at the following expression for the generating function of the restarted process
\bea 
G_{N_{R}}(z)
&=&\frac{\left(1-\text{Pr}\left(N \geq R\right) \right)G_{N_{min}}(z)}{1-\text{Pr}\left(N \geq R\right)G_{R_{min}}(z)}, \label{PGF final result}
\eea 
where $G_{X_{min}}(z)$ is the generating function for the random variable $X_{min}$. The above formula (\ref{PGF final result}), being a simple analogue to Eq. (11) in \cite{RW1}, is extremely useful since it allows one to compute \textit{all the moments}
\bea 
\left\langle N_R^{k}\right\rangle &=&\left(z \frac{\partial}{\partial z}\right)^{k}G_{N_R}(z)\Big|_{z=1^{-}}, \label{PGF moments}
\eea 
and, importantly, also the \textit{probability density function} of $N_R$
\bea
P_{N_R}(n)&=&\text{Pr}(N_R=n)=\frac{G_{N_R}^{(n)}(0)}{n!}, \label{PGF PMF}
\eea
where $\left\langle N_R^{k}\right\rangle$ is the $k$-th moment and $G_{X}^{(n)}(z)$ is the $n$-th derivative of $G_{X}(z)$ with respect to $z$, and $n=0,1,2...$. Note that in discrete set-up,   computation of the first passage time probability mass requires only derivatives of the generating function as seen in \eref{PGF PMF}, and thus is more accessible compared to the continuous time set-up \cite{FPUR}.

\subsection{First passage and survival probability}

Depending on the quantities of interest, survival functions are often computed to infer first passage statistics. In $z$-space, the first passage and survival functions are almost trivially connected. For the random variable $N$, we can write
\bea
&G_{Q_{N}}(z)&\equiv \sum_{n=0}^{\infty}z^{n}Q_{N}(n)=\frac{1-G_{N}(z)}{1-z},\label{PGF survival} 
\eea
where recall $G_N(z)=\sum_{n=0}^{\infty}z^{n}P_{N}(n)$ is the PGF for first passage time density (see \cite{RW1} for derivation). This relation is very general and useful, and holds for any non-negative random variable $X$ e.g., with restarts, we have
\bea
G_{Q_{N_R}}(z)\equiv \sum_{n=0}^{\infty}z^{n}Q_{N_R}(n)=\frac{1-G_{N_R}(z)}{1-z},\label{PGF survival derivation 2-restart}
\eea
where $G_{N_R}(z)$ is given by \eref{PGF def}.

\section{Geometrically distributed restart}
In this section, we consider a specific form of restart time distribution, namely the Geometric distribution. Here, a resetting step number is taken from the following distribution with parameter $p~  (0<p<1)$, 
\bea
P_R(n)=(1-p)^{n-1} p, ~~n\geq 1.
\label{geometric}
\eea
In other words, restart will occur exactly at the $n$-th step with probability $p$, after $n-1$ unsuccessful attempts. 
Notably, this distribution is the discrete analog of the exponential distribution, being a discrete distribution possessing the memory-less property. 

\subsection{Mean}
To derive the mean completion time, we first note that the survival function for the restart can be written as
\bea
Q_R (n) &=& \nn  \text{Pr}(R>n) \\ &=& \sum_{k=n+1}^{\infty} (1-p)^{k-1} p \nn \\
&=& \sum_{k=1}^{\infty} (1-p)^{k+n-1} p \nn \\
&=& (1-p)^{n}\sum_{k=1}^{\infty} (1-p)^{k-1} p \nn \\ &=& (1-p)^{n}. \label{Survival Geo}
\eea 
We now use \eref{Survival Geo}  and \eref{PGF survival} to calculate 
$\left\langle min\left(N,R\right)\right\rangle$, given in \eref{Mean min formula}
\bea
\left\langle min\left(N,R\right)\right\rangle&=& \sum_{n=0}^{\infty} Q_N(n) Q_R(n) \nn \\ 
&=& Q_N(0) + \sum_{n=1}^{\infty} Q_N(n) (1-p)^{n} \nn \\ 
&=& \sum_{n=0}^{\infty} Q_N(n) (1-p)^{n} \nn \\ 
&=& G_{Q_N}(1-p) \nn \\ 
&=& \frac{1-G_{N}(1-p)}{p}. \label{Mean numerator-main}
\eea
Next, following \eref{denominator}, we have
\bea 
\text{Pr}\left(N < R\right)&=&\sum_{n=0}^{\infty}P_{N}(n)\sum_{m=n+1}^{\infty}P_{R}(m) \nn \\
&=& P_{N}(0)+ \sum_{n=1}^{\infty}P_{N}(n) (1-p)^{n} \nn \\
&=&G_N(1-p), \label{mean denominator geometric} \eea

Substituting Eqs. (\ref{Mean numerator-main}) and (\ref{mean denominator geometric}) into the formula for the mean of the restarted process given in \eref{mean FPUR} yields 
\bea 
\langle N_R \rangle &=& \frac{1-G_N(1-p)}{pG_N(1-p)},
\label{mean FPUR geometric}
\eea 
which gives us a simple plug-n-play solution for the mean completion time with the knowledge of the \textbf{underlying first passage time distribution}. We can also rewrite the mean first passage time in term of the PGF of the \textbf{survival function}. This is done by  substituting \eref{PGF survival} into \eref{mean FPUR geometric}
\bea
\langle N_R \rangle &=& \frac{G_{Q_N}(1-p)}{1-pG_{Q_N}(1-p)}, \label{MFPT Survival}
\eea 
which is identical to Eq. (7), derived by Kusmierz \textit{et al} in \cite{Lukasz}.

\subsection{Generating function for the restarted process}

We now turn to the derivation of the PGF of the restarted process under geometric restart. We first compute the PGFs for the conditional random variables $N_{min}$ and $R_{min}$. Following a similar computation as in \cite{RW1}, 
we find
\bea
G_{N_{min}}(z)
&=& \sum_{n=0}^{\infty} P_{N}(n)\frac{\sum_{m=n+1}^{\infty}P_{R}(m)}{\text{Pr}\left(N < R\right)} z^n \nn \\
&=& \sum_{n=0}^{\infty} P_{N}(n)\frac{(1-p)^{n}}{G_N(1-p)} z^n \nn \\
&=& \frac{1}{G_N(1-p)}\sum_{n=0}^{\infty} P_{N}(n)(1-p)^n z^n \nn \\
&=& \frac{G_N(z(1-p))}{G_N(1-p)},
\label{GNmin geometric}
\eea
and
\bea
G_{R_{min}}(z)&=& \sum_{n=0}^{\infty} P_{R}(n)\frac{\sum_{m=n}^{\infty}P_{N}(m)}{\text{Pr}\left(N \geq R\right)} z^n \nn \\
&=& \sum_{n=1}^{\infty} (1-p)^{n-1}p\frac{Q_N(n-1)}{1-G_N(1-p)} z^n \nn \\
&=& \frac{pz}{1-G_N(1-p)}\sum_{n=0}^{\infty} Q_N(n)(1-p)^n z^n \nn \\
&=& \frac{pzG_{Q_N}((1-p)z)}{1-G_N(1-p)} \nn \\
&=& \frac{pz(1-G_N((1-p)z))}{(1-(1-p)z)(1-G_N(1-p))},
\label{GRmin geometric}
\eea
where in the last step we once again used \eref{PGF survival}.
Substituting Eqs. (\ref{GNmin geometric})-(\ref{GRmin geometric}) into \eref{PGF final result} we find
\bea
G_{N_{R}}(z)
&=& \frac{(1-(1-p) z)G_N((1-p) z)}{1-z+pzG_N((1-p)z)}, \label{PGF geometric}
\eea 
from which one can derive the probability mass function of the restarted process by taking the derivatives of the generating function using \eref{PGF PMF}. 
Now substituting (using \eref{PGF survival})
\bea
G_N(z)& = (z-1)G_{Q_{N}}(z)+1
\eea
and 
\bea
G_{{N_R}}(z)=(z-1)G_{Q_{N_R}}(z)+1
\eea
into \eref{PGF geometric} we obtain a relation between the survival functions with and without restart
\bea 
G_{Q_{N_R}}(z)=\frac{G_{Q_N}((1-p)z)}{1-pzG_{Q_N}((1-p)z)},
\label{PGF survival geometric}
\eea 
which is identical to Eq. (5), derived by Kusmierz \textit{et al} in \cite{Lukasz}. \textbf{While \eref{PGF geometric} connects the first passage time distributions, \eref{PGF survival geometric} relates the survival functions.}

\subsection{Criterion for restart to be beneficial}
\label{CV-criterion-geometric}
To derive the criterion, we first observe a first passage time process and turn on an infinitesimal restart probability $p \to 0^+$. If restart has to lower the mean time, it is sufficient enough to check whether $d\langle N_R \rangle/dp|_{p \to 0}<0$, where $\langle N_R \rangle$ is given by \eref{mean FPUR geometric}. A small $p$ expansion of $\langle N_R \rangle$ gives
\bea 
\langle N_R \rangle \approx G_N'(1)+\frac{1}{2} p \left(2 G_N'^2(1)-G_N''(1)\right). \label{Mean around p = 0}
\eea 
Now noting that $G_N'(1)=\langle N \rangle$, $G_N''(1)+G_N'(1)-G_N'(1)^2=Var(N)$ and substituting into \eref{Mean around p = 0}, the criterion can be recast as
\bea
CV^2>1+\frac{1}{\langle N \rangle}
\label{CV crierion}
\eea
where $CV^2=\frac{Var(N)}{\langle N \rangle^2}$ is the squared coefficient of variation of the underlying first passage process. This essentially means that whether restart would favour a completion depends on the underlying first passage time process. Moreover, this criterion is also not sensitive to the entire density, but only to the first two moments of the underlying process. Note that this criterion is slightly different than from the one used in \cite{RW1} where we assumed that process can complete even when the steps for first passage and restart coincide i.e., $N_R=N$ when $N \leq R$ unlike here [see \eref{Renewal eq} in above]. Also see \cite{RW5} where they obtained the same criterion. We refer to a similar criterion that was derived for the continuous stochastic resetting case in \cite{FPUR}.

\section{Applications}
In this section, we revisit a few lattice restarted random walk problems which were studied recently by different groups (and that will be mentioned in context).

\subsection{Simple random walk}
Consider a simple random walker (starting at the origin) in the presence of only one absorbing boundary located at $x$. In this case, the generating function for the underlying process can be easily obtained following steps from \cite{LRW-5} and this reads 
\begin{align}
    G_N(z) =\left(\frac{1-\sqrt{1-z^2}}{z}\right)^{|x|},\quad x\neq 0. \label{Simple random walk first passage}
\end{align}
In fact, it is known that for large $n$, the first passage time density has a power law tail $n^{-3/2}$ \cite{LRW-4,LRW-5,LRW-6}, which is similar to the L\'{e}vy-Smirnov distribution for the first passage time of a Brownian walker in one dimension. This power law trivially leads to a diverging mean first passage time for a RW. However, motion of the walker is restarted to the origin with a probability $p$. This simple restart mechanism can lead to finiteness of the mean first passage time (like the simple diffusion problem in \cite{Restart1} and see the recent review \cite{Restart-review}). 
The mean completion time of a random walker on this geometry reads
\bea 
\langle N_R \rangle=\frac{1}{p}\left[ \left( \frac{1+\sqrt{2p-p^2}}{1-p} \right)^{|x|}-1  \right] ,\label{mean FPUR simple RW}
\eea 
which can be obtained by substituting \eref{Simple random walk first passage} into \eref{mean FPUR geometric}. This result was previously obtained by Riascos \textit{et al} [see Eq. (26)  in \cite{network-2}]. 

\subsection{One sided Sisyphus random walk}
In \cite{RW2}, Montero and Villarroel studied one sided Sisyphus random walk in the presence of geometric restarts. In this dynamics, the walker starts from the origin and makes a deterministic step to the right at each time step. However, the motion is stopped with probability $p$ and the walker is returned to the origin. The process ends as soon as the walker reaches the threshold $a>0$. The first passage time is recorded accordingly.
Since the process is deterministic, for the underlying first passage process, we should have
\bea 
N = a, \label{One direction sisyphus}
\eea
so that $P_N(n)=\delta_{n,a}$, where $\delta_{a,b}$ is the Kronecker delta.
The PGF of this first passage process is given by 
\bea 
G_N(z)=\sum_{n=0}^{\infty} P_{N}(n)z^{n} = z^a.  \label{PGF one sided sisyphus}
\eea 
Substituting the PGF of the underlying process given in \eref{PGF one sided sisyphus} into \eref{PGF geometric} yields
\bea 
G_{N_R}(z)=\frac{(1 - (1 - p) z) (z(1-p))^a}{1 - z (1 - p (z(1-p)^a)}. \label{restart PGF one sided sisyphus 1}
\eea 
Substituting $p\rightarrow1-q$ (keeping the notation as in \cite{RW2}) yields 
\bea 
G_{N_R}(z)=(qz)^a\frac{1 - qz}{1 - z +(1-q)z(qz)^a}, \label{restart PGF one sided sisyphus 2}
\eea 
which is identical to Eq. (16) in \cite{RW2}. Furthermore, substituting the PGF into \eref{mean FPUR geometric} gives us
\bea 
\langle N_R \rangle = \frac{1}{1-q}\left(\frac{1}{q^a}-1\right), \label{mean sisyphus}
\eea 
(again by replacing $p \to 1-q$) which is identical to Eq. (18) in \cite{RW2}. 

\subsection{Two sided Sisyphus random walk}
In a recent preprint \cite{RW3}, Villarroel \textit{et al} extended their study of one sided Sisyphus random walk to a two sided Sisyphus walk resulting in the walker in a 1D confinement (see \cite{interval1,interval2} for such other examples in continuous set-ups). 
The underlying first passage process in the interval $[-b,a]$ can be understood in the following way
\bea 
N = \begin{cases}
b & \text{prob.} \quad 1-\rho\\
a & \text{prob.} \quad \rho
\end{cases}. \label{Sisyphus underlying process}
\eea
This means that the walker deterministically moves to the right with probability $\rho$ and to the left with the complementary probability. The process ends as soon as the walker reaches one of the boundaries. The PGF of $N$ is then given by
\bea 
G_{N}(z) = \sum_{n=0}^{\infty} P_{N}(n)z^{n} = z^b(1-\rho)+z^a \rho.  \label{PGF Sisyphus}
\eea 
Substituting the PGF of the underlying process given in \eref{PGF Sisyphus} into \eref{PGF geometric}, and substituting $p\rightarrow 1-q$ yields the generating function for the entire process
\begin{align}
G_{N_R}(z)=\frac{(1-qz)((qz)^a\rho +(qz)^b(1-\rho))}{1-z+(1-q)z((qz)^a\rho +(qz)^b(1-\rho))}. \label{PGF sisyphus}
\end{align}
This is a useful result since it helps one to derive not only the mean but also the higher order moments. For a symmetric box, \eref{PGF sisyphus} reduces to 
\bea 
G_{N_R}(z)=(qz)^a\frac{1 - qz}{1 - z +(1-q)z(qz)^a}, \label{PGF symmetric sisyphus}
\eea 
which is identical to \eref{restart PGF one sided sisyphus 2}, as it should [also see Eq. (51) in \cite{RW3}]. The results are identical since a walk directed to the left boundary and a walk directed to the right boundary are identically distributed for the case of a symmetric box. Thus, walks in opposite directions can be considered as different realizations of the same process. 

Substituting the PGF given in \eref{PGF Sisyphus} into \eref{mean FPUR geometric}, and $p\rightarrow 1-q$, yields
the mean escape time for the walker from the interval in the presence of restart 
\bea 
\langle N_R \rangle = \frac{ 1-(\rho q^a +(1-\rho)q^b)}{(1-q)(\rho q^a +(1-\rho)q^b)}. \label{mean FPUR Sisyphus}
\eea
In the case of symmetric box, the mean first passage reduces to
\bea 
\langle N_R \rangle = \frac{1}{1-q}\left(\frac{1}{q^a}-1\right).\label{mean FPUR symmetric Sisyphus}
\eea 
which is identical to \eref{mean sisyphus}. In \cite{RW3}, the authors introduced $l$ as the average number of steps the walker takes until resetting. Translating to our language $\langle R \rangle =l$, and also $l=\frac{1}{1-p}=\frac{1}{q}$. 
Substituting this into \eref{mean FPUR symmetric Sisyphus}, we get
\bea 
\langle N_R \rangle = l\frac{l^a-1}{l-1}, \label{mean FPUR symmetric Sisyphus 2}
\eea 
which was also derived in \cite{RW3} [see Eq. (55)]. It is important to note that our results hold for asymmetric interval, and thus are more general. 

\section{Examples of Other restart protocols}
In this section, we present exact results for the mean completion time of a generic first passage process (discrete space and time) under discrete restart protocols which are not necessarily memoryless. In other words, unlike geometric distribution, they may not be often thought as a probabilistic way. In fact, the protocols we choose can also come from a heavy-tailed distribution. \textbf{We show how our formalism is robust to all these changes and finally the results are given in terms of the statistical metrics of the underlying first passage process.}

\subsection{Sharp restart}
Consider now a strategy when restart events always take place after a fixed number of steps. This is often known as sharp, periodic or deterministic restart protocol (see \cite{PalJphysA,FPUR} for more details of this strategy). Since the resetting takes place always after a fixed period, the density can be written as
\bea 
P_R(n)=\delta_{n,r}=\begin{cases}
0, & n\neq r\\
1, & n=r
\end{cases}, \label{Sharp distribution}
\eea 
where $\delta_{n,r}$ is the Kronecker delta. So, we will refer to this as a distribution with restart step or period $r$. We have studied this set-up in the preceding work \cite{RW1}. Here, we recall the results for brevity. 
For sharp restart, we have
\bea
\text{Pr}\left(N<R\right)=\text{Pr}\left(N<r\right)=\sum_{n=0}^{r-1}P_{N}(n).
\label{sharp-num}
\eea
Furthermore, we find 
\bea
\langle \text{min} (N,R) \rangle &=& \sum_{n=0}^{\infty} Q_N(n) Q_R(n) \nn \\
&=& \sum_{n=0}^{\infty} Q_N(n) \sum_{m=n+1}^{\infty} P_R(m) \nn \\
&=& \sum_{n=0}^{\infty} Q_N(n) \sum_{m=n+1}^{\infty} \delta _{m,r} \nn \\
&=& \sum_{n=0}^{r-1} Q_N(n). \label{Min sharp}
\eea 
Substituting \eref{sharp-num} and \eref{Min sharp} into \eref{mean FPUR} we get the following expression for the mean completion time
\bea 
\langle N_R \rangle = \frac{ \sum_{n=0}^{r-1} Q_N(n)}{\sum_{n=0}^{r-1}P_{N}(n)}, \label{mean sharp}
\eea 
which is given in terms of the survival and first passage of the underlying process.

\subsection{Poisson restart}
Here we consider the  restart steps to be drawn from the Poisson distribution namely
\bea 
P_R(n) = \frac{\lambda^{n-1} e^{-\lambda}}{(n-1)!}, n=1,2,3,...
\eea 
where $\lambda$ is a positive rate parameter.
Here, we use a Poisson distribution with rate parameter $\lambda$, shifted by 1, so the support of the restart distribution is strictly positive, i.e., avoiding restart at 0.
Under this resetting protocol, we get
\bea 
\text{Pr}(N<R) &=& \sum _{n=0}^{\infty} P_N(n) \sum _{m=n+1}^{\infty} P_R(m) \nn \\
&=& P_N(0) + \sum _{n=1}^{\infty} P_N(n) \sum _{m=n+1}^{\infty} P_R(m) \nn \\ &=& P_N(0)+ \sum _{n=1}^{\infty} P_N(n) \left( 1 - \sum _{m=1}^{n} P_R(m) \right ) \nn \\
&=& \sum _{n=0}^{\infty} P_N(n) - \sum _{n=1}^{\infty} P_N(n) \sum _{m=1}^{n} P_R(m) \nn \\
&=& 1-\sum _{n=1}^{\infty} P_N(n) \sum _{m=1}^{n} P_R(m) \nn \\
&=& 1-\sum _{n=1}^{\infty} \sum _{m=1}^{n} P_N(n) P_R(m) \nn \\
&=& 1-\sum _{m=1}^{\infty} \sum _{n=m}^{\infty} P_N(n) P_R(m) \nn \\
&=& 1-\sum _{m=1}^{\infty} P_R(m) \sum _{n=m}^{\infty} P_N(n) \nn \\
&=& 1-\sum _{m=1}^{\infty} P_R(m) Q_N(m-1) \nn \\ 
&=& 1-\sum _{m=0}^{\infty} P_R(m+1) Q_N(m) \nn \\ 
&=& 1-e^{-\lambda}\sum _{m=0}^{\infty} \frac{\lambda^m}{m!} Q_N(m), \label{Pr poisson} 
\eea
and
\bea 
\langle \text{min} (N,R) \rangle &=& \sum _{n=0}^{\infty} Q_N(n) Q_R(n) \nn \\
&=& Q_N(0) + \sum _{n=1}^{\infty} Q_N(n) Q_R(n) \nn \\
&=& Q_N(0) + \sum _{n=1}^{\infty} Q_N(n) \left( 1 - \sum _{m=1}^{n} P_R(m) \right) \nn \\
&=& \sum _{n=0}^{\infty} Q_N(n) - \sum _{n=1}^{\infty} Q_N(n) \sum _{m=1}^{n} P_R(m) \nn \\
&=& \langle N \rangle - \sum _{n=1}^{\infty} Q_N(n) \sum _{m=1}^{n} \frac{\lambda^{m-1} e^{-\lambda}}{(m-1)!} \nn \\
&=& \langle N \rangle - \sum _{n=1}^{\infty} Q_N(n) \sum _{m=0}^{n-1} \frac{\lambda^{m} e^{-\lambda}}{m!} \nn \\
&=& \langle N \rangle - e^{-\lambda} \sum _{n=1}^{\infty} Q_N(n) \sum _{m=0}^{n-1} \frac{\lambda^{m}}{m!}. \label{Min poisson}
\eea 
Substituting \eref{Pr poisson} and \eref{Min poisson} into \eref{mean FPUR} we get the following formula for the mean first passage time under Poissonian resetting
\bea
\langle N_R \rangle = \frac{\langle N \rangle - e^{-\lambda} \sum _{n=1}^{\infty} Q_N(n) \sum _{m=0}^{n-1} \frac{\lambda^{m}}{m!}}{1-e^{-\lambda}\sum _{m=0}^{\infty} \frac{\lambda^m}{m!} Q_N(m)}, \label{mean poisson}
\eea 
which is given in terms of the survival function of the underlying first passage time process. 

\subsection{Zeta restart}
We now consider restart steps which are taken from Zeta distribution
\bea 
P_R(n) = \frac{n^{-s}}{\zeta(s)}, n=1,2,3,...
\eea 
where $s$ is a positive integer, and $\zeta(s)$ is the Zeta function, defined as
\bea 
\zeta (s) = \sum _{m=1}^{\infty} \frac{1}{m^s}, Re(s)>1.
\eea 
The Zeta distribution is a heavy-tailed distribution. 
Under this resetting protocol, following similar steps to the Poisson case, we get
\bea 
\text{Pr}(N<R) &=& \sum _{n=0}^{\infty} P_N(n) \sum _{m=n+1}^{\infty} P_R(m) \nn \\
&=& P_N(0) + \sum _{n=1}^{\infty} P_N(n) \sum _{m=n+1}^{\infty} P_R(m) \nn \\ &=& P_N(0)+ \sum _{n=1}^{\infty} P_N(n) \left( 1 - \sum _{m=1}^{n} P_R(m) \right ) \nn \\
&=& \sum _{n=0}^{\infty} P_N(n) - \sum _{n=1}^{\infty} P_N(n) \sum _{m=1}^{n} P_R(m) \nn \\
&=& 1-\sum _{n=1}^{\infty} P_N(n) \sum _{m=1}^{n} P_R(m) \nn \\
&=& 1-\sum _{n=1}^{\infty} \sum _{m=1}^{n} P_N(n) P_R(m) \nn \\
&=& 1-\sum _{m=1}^{\infty} \sum _{n=m}^{\infty} P_N(n) P_R(m) \nn \\
&=& 1-\sum _{m=1}^{\infty} P_R(m) \sum _{n=m}^{\infty} P_N(n) \nn \\
&=& 1-\sum _{m=1}^{\infty} P_R(m) Q_N(m-1) \nn \\ 
&=& 1-\frac{1}{\zeta(s)}\sum _{m=1}^{\infty} Q_N(m-1)m^{-s}, \label{Pr Zeta} 
\eea
and
\bea 
\langle \text{min} (N,R) \rangle &=& \sum _{n=0}^{\infty} Q_N(n) Q_R(n) \nn \\
&=& Q_N(0) + \sum _{n=1}^{\infty} Q_N(n) Q_R(n) \nn \\
&=& Q_N(0) + \sum _{n=1}^{\infty} Q_N(n) \left( 1 - \sum _{m=1}^{n} P_R(m) \right) \nn \\
&=& \sum _{n=0}^{\infty} Q_N(n) - \sum _{n=1}^{\infty} Q_N(n) \sum _{m=1}^{n} P_R(m) \nn \\
&=& \langle N \rangle - \sum _{n=1}^{\infty} Q_N(n) \sum _{m=1}^{n} \frac{m^{-s}}{\zeta(s)} \nn \\
&=& \langle N \rangle - \frac{1}{\zeta(s)} \sum _{n=1}^{\infty} Q_N(n) \sum _{m=1}^{n} m^{-s} \nn \\
&=& \langle N \rangle - \frac{1}{\zeta(s)} \sum _{n=1}^{\infty} Q_N(n) H_{n,s}, \label{Min Zeta}
\eea 
where $H_{n,s}$ is the generalized harmonic number of order $s$ of $n$.
Substituting \eref{Pr Zeta} and \eref{Min Zeta} into \eref{mean FPUR} we get the following formula for the mean first passage time under Zeta distributed resetting
\bea
\langle N_R \rangle = \frac{\langle N \rangle - \frac{1}{\zeta(s)} \sum _{n=1}^{\infty} Q_N(n) H_{n,s}}{1-\frac{1}{\zeta(s)}\sum _{m=1}^{\infty} Q_N(m-1)m^{-s}},
\label{mean Zeta}
\eea 
which is given in terms of the survival function of the underlying process.

\section{Concluding perspective}
In this note, we presented a simple, and completely general recipe to study statistical properties of discrete first passage processes under arbitrary discrete restart steps. The stages of the computation are as follows:

\begin{itemize}
    \item \textbf{Survival functions:} compute the survival functions $Q_N(n)$ and $Q_R(n)$ from the given first passage and restart time distributions namely $P_N(n)$ and $P_R(n)$ respectively [use Eqs. (\ref{Survival N}) and (\ref{Survival R})].
    \item \textbf{Mean completion time under restart:} next, compute $\text{Pr}(N<R)$ and its complementary probability $\text{Pr}(N \geq R)$ using \eref{denominator}. The mean completion time can then be obtained from \eref{mean FPUR}.
    \item \textbf{Generating function of the restarted first passage process:} compute the generating functions for the random variable $N_{min}$ and $R_{min}$ and plug in \eref{PGF final result} to derive the generating function for the restarted process from which all the moments can be derived. In some cases, the first passage time density of the compound process can be obtained making use of \eref{PGF PMF}. This was demonstrated in \cite{RW1}. \\
\end{itemize}

\noindent
The three-step algorithm prescribed above gives a systematic way to compute first passage quantities 
and investigate the effects of different restart mechanisms.
Since the formalism holds for any
first passage process (Markov or non-Markov) under any restart mechanism (Markov or non-Markov), we anticipate that the method would be useful to develop synergy between discrete stochastic process and resetting which are much less studied compared to the continuous space-time processes. \\

\noindent
\textbf{Numerical treatment.---} 
Finally, it is worth emphasizing that the machinery presented above also works to treat the problems numerically without having to solve the exact dynamics of the stochastic process of interest. What one needs is two datasets: one is for the underlying first passage time (which is known for many problems numerically) and the other for resetting (which is an external input). The above-mentioned three-step recipe can then be implemented numerically, and relevant quantities may be computed. To this end, a supplemented \textbf{MATHEMATICA} file will be available on the Notebook Archive.

\begin{acknowledgments}
AP gratefully acknowledges support from the Raymond and  Beverly  Sackler  Post-Doctoral  Scholarship and the Ratner Center for Single Molecule Science at  Tel-Aviv University. ...
\end{acknowledgments}

\end{document}